\documentclass[twocolumn,floatfix,aps,superscriptaddress,nofootinbib]{revtex4-1}
\usepackage{graphicx}
\usepackage{amsmath}
\usepackage{amssymb}
\usepackage{bm}

\usepackage{color}

\begin{document}

\title{Superconductivity provides access to the chiral magnetic effect\\
of an unpaired Weyl cone}
\author{T. E. O'Brien}
\affiliation{Instituut-Lorentz, Universiteit Leiden, P.O. Box 9506, 2300 RA Leiden, The Netherlands}
\author{C. W. J. Beenakker}
\affiliation{Instituut-Lorentz, Universiteit Leiden, P.O. Box 9506, 2300 RA Leiden, The Netherlands}
\author{\.{I}. Adagideli}
\affiliation{Faculty of Engineering and Natural Sciences, Sabanci University, Orhanli-Tuzla, Istanbul, Turkey}
\affiliation{Instituut-Lorentz, Universiteit Leiden, P.O. Box 9506, 2300 RA Leiden, The Netherlands}
\date{February 2017}
\begin{abstract}
The massless fermions of a Weyl semimetal come in two species of opposite chirality, in two cones of the band structure. As a consequence, the current $j$ induced in one Weyl cone by a magnetic field $B$ (the chiral magnetic effect, CME) is cancelled in equilibrium by an opposite current in the other cone. Here we show that superconductivity offers a way to avoid this cancellation, by means of a flux bias that gaps out a Weyl cone jointly with its particle-hole conjugate. The remaining gapless Weyl cone and its particle-hole conjugate represent a single fermionic species, with renormalized charge $e^\ast$ and a single chirality $\pm$ set by the sign of the flux bias. As a consequence, the CME is no longer cancelled in equilibrium but appears as a supercurrent response $\partial j/\partial B=\pm(e^\ast e/h^2)\mu$ along the magnetic field at chemical potential $\mu$. 
\end{abstract}
\maketitle

\emph{Introduction ---}
Massless spin-$1/2$ particles, socalled Weyl fermions, remain unobserved as elementary particles, but they have now been realized as quasiparticles in a variety of crystals known as Weyl semimetals \cite{Ber15,Ciu15,Jia16,Bur16,Rao16}. Weyl fermions appear in pairs of left-handed and right-handed chirality, occupying a pair of cones in the Brillouin zone. The pairing is enforced by the chiral anomaly \cite{Nie83}: A magnetic field induces a current of electrons in a Weyl cone, flowing along the field lines in the chiral zeroth Landau level. The current in the Weyl cone of one chirality has to be canceled by a current in the Weyl cone of opposite chirality, to ensure zero net current in equilibrium. The generation of an electrical current density $\bm{j}$ along an applied magnetic field $\bm{B}$, the socalled chiral magnetic effect (CME) \cite{Kha14,Bur15}, has been observed as a dynamic, nonequilibrium phenomenon \cite{Kim13,Xio15,Hua15,Li15,Li16} --- but it cannot be realised in equilibrium because of the fermion doubling \cite{Vaz13,Zho13,Che13,Gos13,Bas13,Cha15,Ma15,Ala16,Zho16a,Bai16a,Zub16}.

\begin{figure}[tb]
\centerline{\includegraphics[width=0.9\linewidth]{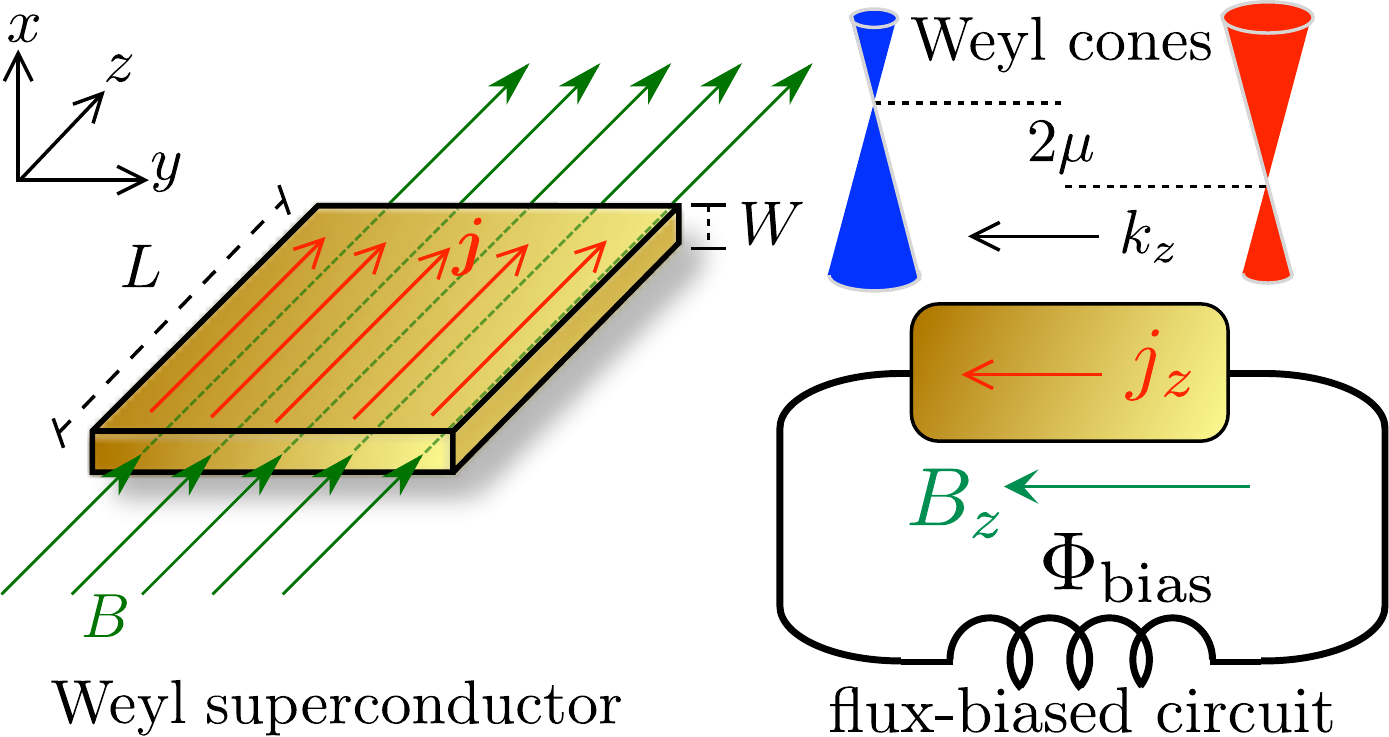}}
\caption{Left panel: Slab of a Weyl superconductor subject to a magnetic field $B$ in the plane of the slab (thickness $W$ less than the London penetration depth). The equilibrium chiral magnetic effect manifests itself as a current response $\partial j/\partial B=\pm\kappa (e/h)^2\mu$ along the field lines, with $\kappa$ a charge renormalization factor and $\mu$ the equilibrium chemical potential. The right panel shows the flux-biased measurement circuit and the charge-conjugate pair of Weyl cones responsible for the effect, of a single chirality $\pm$ determined by the sign of the flux bias.
}
\label{fig_layout}
\end{figure}

Here we present a method by which single-cone physics may be accessed in a superconducting Weyl semimetal, allowing for observation of the CME in equilibrium. The geometry is shown in Fig.\ \ref{fig_layout}. Application of a flux bias gaps out all but a single particle-hole conjugate pair of Weyl cones, of a single chirality $\pm$ set by the sign of the flux bias. At nonzero chemical potential $\mu$, one of the two Weyl points sinks in the Cooper pair sea, the chiral anomaly is no longer cancelled, and we find an equilibrium response $\partial j/\partial B=\pm(e^\ast e/h^2) \mu$, with $e^\ast$ the charge expectation value at the Weyl point.

We stress that the CME in a superconductor is not in violation of thermodynamics, which only demands a vanishing heat current in equilibrium. Indeed, in previous work on magnetically induced currents \cite{Lev85,Naz86,Min13} it was shown that the fundamental principles of Onsager symmetry and gauge invariance forbid a linear relation between $\bm{j}$ and $\bm{B}$ in equilibrium. However, in a superconductor the gauge symmetry is broken at a fixed phase of the order parameter, opening the door for the CME. 

\emph{Pathway to single-cone physics ---}
We first explain the mechanism by which a superconductor provides access to single-cone physics. A pair of Weyl cones at momenta $\pm{\bm k}_0$ of opposite chirality has Hamiltonian \cite{Vol03}
\begin{equation}
{\cal H}=\tfrac{1}{2}v_{\rm F}\textstyle{\sum_{\bm{k}}}\bigl[\psi^\dagger_{\bm k}(\bm{k}-\bm{k}_0)\cdot\bm{\sigma}\psi_{\bm k}^{\vphantom{\dagger}}-\phi_{\bm{k}}^\dagger(\bm{k}+\bm{k}_0)\cdot\bm{\sigma}\phi_{\bm{k}}^{\vphantom{\dagger}}\bigr],\label{HLRdef}
\end{equation}
where $\bm{k}\cdot\bm{\sigma}=k_x\sigma_x+k_y\sigma_y+k_z\sigma_z$ is the sum over Pauli matrices acting on the spinor operators $\psi$ and $\phi$ of left-handed and right-handed Weyl fermions. The Fermi velocity is $v_{\rm F}$ and we set $\hbar\equiv 1$ (but keep $h$ in the formula for the CME).

If ${\cal H}$ would be the Bogoliubov-De Gennes (BdG) Hamiltonian of a superconductor, particle-hole symmetry would require that $\phi_{\bm{k}}=\sigma_y\psi^\dagger_{-\bm{k}}$. With the help of the matrix identity $\sigma_y\sigma_\alpha\sigma_y=-\sigma_\alpha^{\ast}$ and the anticommutator $\psi\sigma_\alpha^\ast\psi^\dagger=-\psi^\dagger\sigma_\alpha\psi$ we rewrite Eq.\ \eqref{HLRdef} as
\begin{align}
{\cal H}&=\tfrac{1}{2}v_{\rm F}\textstyle{\sum_{\bm{k}}}\bigl[\psi^\dagger_{\bm k}(\bm{k}-\bm{k}_0)\cdot\bm{\sigma}\psi_{\bm k}^{\vphantom{\dagger}}-\psi_{\bm{-k}}^\dagger(\bm{k}+\bm{k}_0)\cdot\bm{\sigma}\psi_{-\bm{k}}^{\vphantom{\dagger}}\bigr]\nonumber\\
&=v_{\rm F}\textstyle{\sum_{\bm{k}}}\psi^\dagger_{\bm k}(\bm{k}-\bm{k}_0)\cdot\bm{\sigma}\psi_{\bm k}^{\vphantom{\dagger}},\label{Hsingle}
\end{align}
producing a single-cone Hamiltonian. If we then, hypothetically, impose a magnetic field $\bm{B}=\nabla\times\bm{A}$ via $\bm{k}\mapsto\bm{k}-e\bm{A}$, the zeroth Landau level carries a current density $\bm{j}=(e/h)^2\mu\bm{B}$ in an energy interval $\mu$. This is the chiral anomaly of an unpaired Weyl cone \cite{Nie83}.

\emph{Model Hamiltonian of a Weyl superconductor ---}
As a minimal model for single-cone physics we consider the BdG Hamiltonian \cite{Bai16}
\begin{subequations}
\label{HBdGdef}
\begin{align}
{\cal H}={}&\textstyle{\sum_{\bm{k}}}\Psi^\dagger_{\bm{k}}H(\bm{k})\Psi_{\bm{k}}^{\vphantom{\dagger}},\;\;\Psi_{\bm{k}}=\bigl(\psi_{\bm{k}},\sigma_y\psi^\dagger_{-\bm{k}}\bigr),\\
H(\bm{k})={}&\begin{pmatrix}
H_0(\bm{k}-e\bm{A})&\Delta_0\\
\Delta_0^\ast&-\sigma_y H_0^\ast(-\bm{k}-e\bm{A})\sigma_y
\end{pmatrix},
\\
H_0(\bm{k})={}&\textstyle{\sum_\alpha}\tau_z \sigma_\alpha\sin k_\alpha +\tau_0(\beta\sigma_z -\mu\sigma_0)+m_{\bm k}\tau_x\sigma_0,\nonumber\\
m_{\bm k} ={}& m_0 + \textstyle{\sum_\alpha}  (1-\cos k_\alpha).\label{eq:mass_term}
\end{align}
\end{subequations}
This is a tight-binding model on a simple cubic lattice (lattice constant $a_0\equiv 1$, nearest-neighbor hopping energy $t_0\equiv 1$, electron charge $+e$). The Pauli matrices $\tau_\alpha$ and $\sigma_\alpha$, with $\alpha\in\{x,y,z\}$, act respectively on the orbital and spin degree of freedom. (The corresponding unit matrices are $\tau_0$ and $\sigma_0$.) Time-reversal symmetry is broken by a magnetization $\beta$ in the $z$-direction, $\mu$ is the chemical potential, $\bm{A}$ the vector potential, and $\Delta_0$ is the $s$-wave pair potential.

The single-electron Hamiltonian $H_0$ in the upper-left block of $H$ is the four-band model \cite{Vaz13,Yan11} of a Weyl semimetal formed from a topological insulator in the Bi$_2$Se$_3$ family, layered in the $x$--$y$ plane. For a small mass term $m_0<\beta$ it has a pair of Weyl cones centered at $\bigl(0,0,\pm \sqrt{\beta^2-m_0^2}\bigr)$, displaced in the $k_z$-direction by the magnetization. (We retain inversion symmetry, so the Weyl points line up at the same energy.) A coupling of this pair of electron Weyl cones to the pair of particle-hole conjugate Weyl cones in the lower-right block of $H$ is introduced by the pair potential, which may be realized by alternating the layers of topological insulator with a conventional BCS superconductor \cite{Men12,Bed15}. (Intrinsic superconducting order in a doped Weyl semimetal, with more unconventional pair potentials, is an alternative possibility \cite{Cho12,Shi13,Wei14,Hos14,Kob15,Zho16,Wan16,Has16,Ali16,Far16}.) The superconductor does not gap out the Weyl cones if $\Delta_0<\sqrt{\beta^2-m_0^2}$.

\begin{figure}[tb]
\centerline{\includegraphics[width=\linewidth]{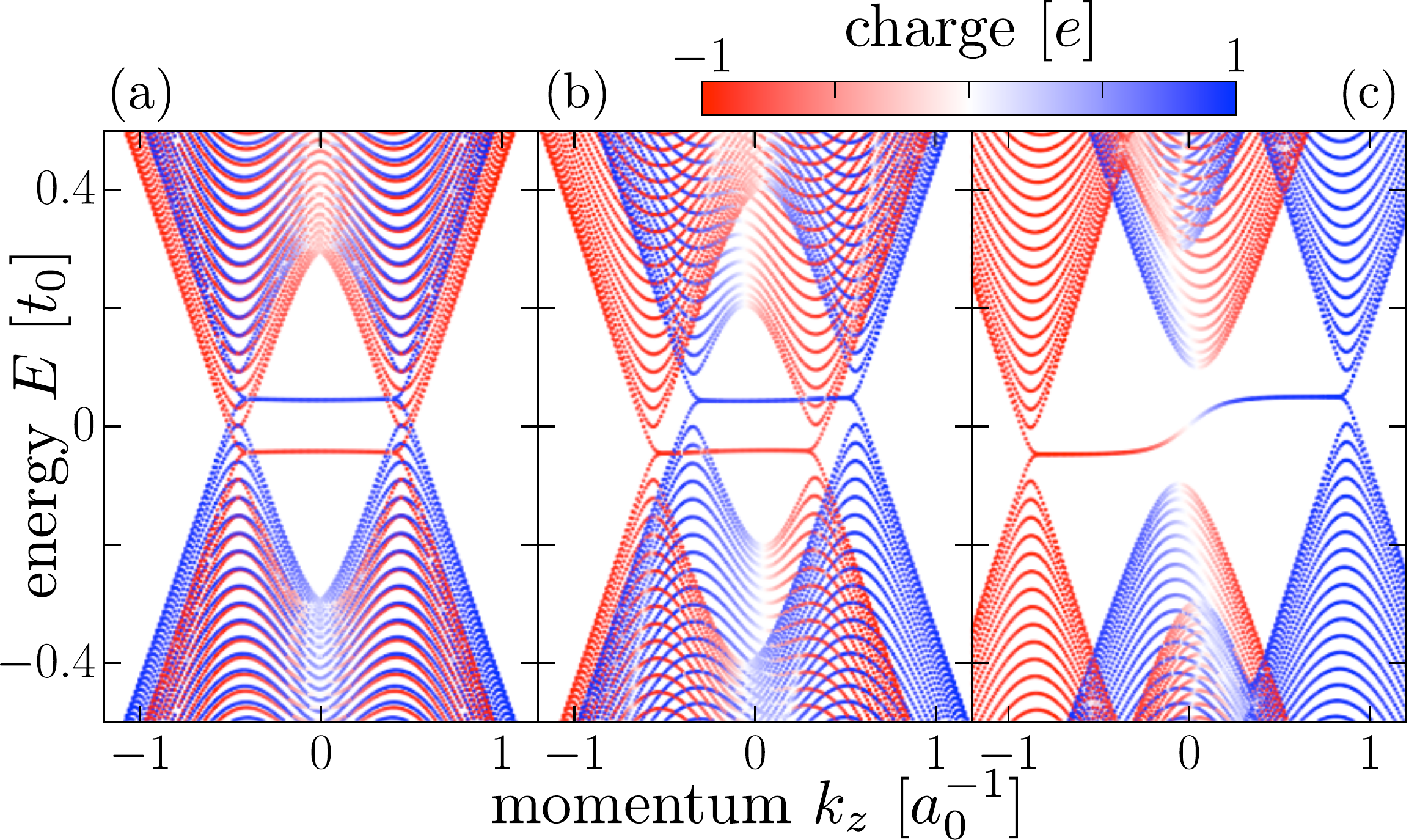}}
\caption{Effect of a flux bias on the band structure of a Weyl superconductor. The plots are calculated from the Hamiltonian \eqref{HBdGdef} in the slab geometry of Fig.\ \ref{fig_layout} (parameters: $m_0=0$, $\Delta_0=0.2$, $\beta=0.5$, $\mu=-0.05$, $k_y=0$, $W=100$, $B_z=0$). The color scale indicates the charge expectation value, to distinguish electron-like and hole-like cones. As the flux bias is increased from $\Lambda=0$ in panel (a), to $\Lambda=0.1$ and $0.4$ in panels (b) and (c), one electron-hole pair of Weyl cones merges and is gapped by the pair potential. What remains in panel (c) is a single pair of charge-conjugate Weyl cones, connected by a surface Fermi arc. This is the phase that supports a chiral magnetic effect in equilibrium.
}
\label{fig:transform_fig}
\end{figure}

\emph{Flux bias into the single-cone regime ---}
As explained by Meng and Balents \cite{Men12}, a Weyl superconductor has topologically distinct phases characterized by the number ${\cal N}\in\{2,1,0\}$ of ungapped particle-hole conjugate pairs of Weyl cones. We propose to tune through the phase transitions in an externally controllable way by means of a flux bias, as shown in the circuit of Fig.\ \ref{fig_layout}. For a real $\Delta_0>0$ the flux bias $\Phi_{\rm bias}$ enters in the Hamiltonian via the vector potential component $A_z=\Phi_{\rm bias}/L\equiv \Lambda/e$. The $\Phi_{\rm bias}$-dependent band structure is shown in Fig.\ \ref{fig:transform_fig}, calculated \cite{kwant} in a slab geometry with hard-wall boundaries at $x=\pm W/2$ and periodic boundary conditions at $y=\pm W'/2$ (sending $W'\rightarrow\infty$).

The two pairs of particle-hole conjugate Weyl cones are centered at $ (0,0,K_\pm)$ and $(0,0,-K_\pm)$, with
\begin{equation}
K_\pm^2=\bigl(\sqrt{\beta^2-m_0^2}\pm\Lambda\bigr)^2-\Delta_0^2.\label{Knudef}
\end{equation}
We have assumed $\Lambda$, $K_\pm\ll 1$, so the Weyl cones are near the center of the Brillouin zone. A cone is gapped when $K_\pm$ becomes imaginary, hence the ${\cal N}=1$ phase is entered with increasing $\Lambda>0$ when
\begin{equation}
\sqrt{\beta^2-m_0^2}+\Lambda>\Delta_0>\bigl|\sqrt{\beta^2-m_0^2}-\Lambda\bigr|.
\label{eq:single_cone_phase}
\end{equation}
This is the regime in which we can observe the CME of an unpaired Weyl cone, as we will show in the following.

\begin{figure*}[tb]
\centerline{\includegraphics[width=0.8\linewidth]{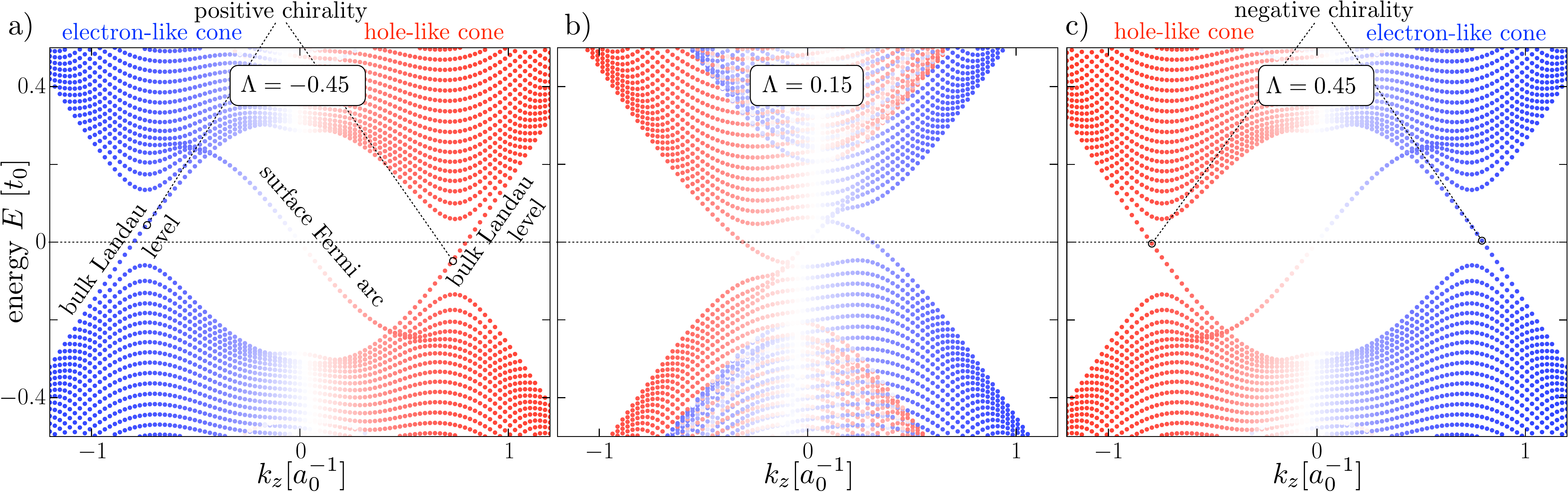}}
\caption{
Chirality switch of a pair of charge-conjugate Weyl cones, induced by a sign change of the flux bias $\Lambda=-0.45$, 0.15, and 0.45 in panels a, b, and c, respectively. All other parameters are the same in each panel: $m_0=0$, $\Delta_0=0.6$, $\beta=0.5$, $W=100$, $k_y=0$, $\mu=-0.05$, and $B_z=0.001\,a_0^{-2}h/e$. The charge color scale of the band structure is as in Fig.\ \ref{fig:transform_fig}. Particles in the zeroth Landau level propagate through the bulk in {\em the same direction} both in the electron-like cone and in the hole-like cone, as determined by the chirality $\chi=-{\rm sign}\,\Lambda$ \cite{note0}. A net charge current appears in equilibrium because $\mu<0$, so there is an excess of electron-like states at $E>0$. [States at $E<0$ do not contribute to the equilibrium current \eqref{Izdef}.] The particle current is cancelled by the Fermi arc that connects the charge-conjugate Weyl cones. The Fermi arc carries an approximately neutral current, hence the charge current in the chiral Landau level is not much affected by the counterflow of particles in the Fermi arc.
}
\label{fig:surface_band}
\end{figure*}

\emph{Magnetic response of a unpaired Weyl cone ---}
We assume that the slab is thinner than the London penetration depth, so that we can impose an unscreened magnetic field $B_z$ in the $z$-direction \cite{vortex}. The vector potential including the flux bias is $\bm{A}=(0,xB_z ,\Lambda/e)$. To explain in the simplest terms how single-cone physics emerges we linearize in $\bm{k}$ and $\bm{A}$ and set $m_0=0$, so the mass term $m_{\bm k}$ can be ignored. (All nonlinearities will be fully included later on \cite{supplemental}.)

The Hamiltonian \eqref{HBdGdef} is approximately block-diagonalized by the Bogoliubov transformation
\begin{equation}
\begin{split}
&\tilde{\psi}_{\bm k}=\cos(\theta_k/2) \psi_{\bm k}^{\vphantom{\dagger}}+i\sin(\theta_k/2) \tau_z\sigma_x\psi_{-{\bm k}}^\dagger,\\
&\tilde{H}=U^{\dagger} HU,\;\;U=\exp\left(\tfrac{1}{2}i\theta_k\nu_y\tau_z\sigma_z\right),
\end{split}
\label{tildepsiHdef}
\end{equation}
where the Pauli matrix $\nu_\alpha$ acts on the particle-hole degree of freedom. If we choose the $k_z$-dependent angle $\theta_k$ such that
\begin{equation}
\begin{split}
&\cos\theta_k=-(\sin k_z)/\Delta_k,\;\;\sin\theta_k=\Delta_0/\Delta_k,\\
&\Delta_k=\sqrt{\Delta_0^2+\sin^2 k_z},
\end{split}\label{thetadef}
\end{equation}
the gapless particle-hole conjugate Weyl points at $k_z^2= K_+^2\approx 2\Delta_0(\beta+\Lambda-\Delta_0)\ll 1$ are predominantly contained in the $(\nu,\tau)=(-,-)$ block of $\tilde{H}$. Projection onto this block gives the low-energy Hamiltonian
\begin{equation}
\tilde{\cal H}=\textstyle{\sum_{\bm{k}}} \tilde\psi_{\bm{k}}^\dagger\bigl[\textstyle{\sum_\alpha}v_\alpha (\delta k_\alpha-q_\alpha A_\alpha)\sigma_\alpha -q_0\mu \sigma_0\bigr] \tilde{\psi}_{\bm{k}},\label{tildecalHdef}
\end{equation}
where $\bm{k}=(0,0,K_+)+\delta\bm{k}$, $\bm{v}=(1,1,-\kappa)$, $q_0=\kappa$,\\
$\bm{q}=(\kappa e,\kappa e,e/\kappa)$,  and
\begin{equation}
\kappa\approx K_+/\sqrt{\Delta_0^2+K_+^2}=\sqrt{1-\Delta_0^2/(\beta+\Lambda)^2}.\label{kappadef}
\end{equation}

Eq.\ \eqref{tildecalHdef} represents a single-cone Hamiltonian of the form \eqref{Hsingle}, with a renormalized velocity $v_\alpha$ and charge $q_\alpha$. As a consequence, the CME formula for the equilibrium current density $j_z$ is renormalized into \cite{supplementalC}
\begin{equation}
\frac{\partial j_z}{\partial B_z}=\frac{q_y q_z}{h^2}q_0\mu=\frac{e^\ast e}{h^2}\mu,\;\;e^\ast=\kappa e . \label{jybulk}
\end{equation}
The renormalization of $\bm{v}$ does not enter because the CME is independent of the Fermi velocity. One can understand why the product $e^\ast e$ appears rather than the more intuitive $(e^\ast)^2$, by noticing that $\partial j_z/\partial B_z$ changes sign upon inversion of the momentum --- hence only odd powers of $\kappa\propto K_+$ are permitted.

\emph{Consistency of a nonzero equilibrium electrical current and vanishing particle current ---}
For thermodynamic consistency, to avoid heat transport at zero temperature, the CME should not produce a particle current in the superconductor. The flow of charge $e^\ast$ particles in the $z$-direction should therefore be cancelled by a charge-neutral counterflow. This counterflow is provided by the surface Fermi arc, as illustrated in Fig.\ \ref{fig:surface_band}. The Fermi arc is the band of surface states connecting the Weyl cones \cite{Wan11,Hal14}, to ensure that the chirality of the zeroth Landau level does not produce an excess number of left-movers over right-movers. In a Weyl superconductor one can distinguish a trivial or nontrivial connectivity, depending on whether the Fermi arc connects cones of the same or of opposite charge \cite{Bai16,YLi15}. Here the connectivity is necessarily nontrivial, because there is only a single pair of charge-conjugate Weyl cones. As a consequence, the Fermi arc is approximately charge neutral near the Fermi level (near $E=0$), so it can cancel the particle current without cancelling the charge current \cite{note1,note2}.

\begin{figure}[tb]
\centerline{\includegraphics[width=1\linewidth]{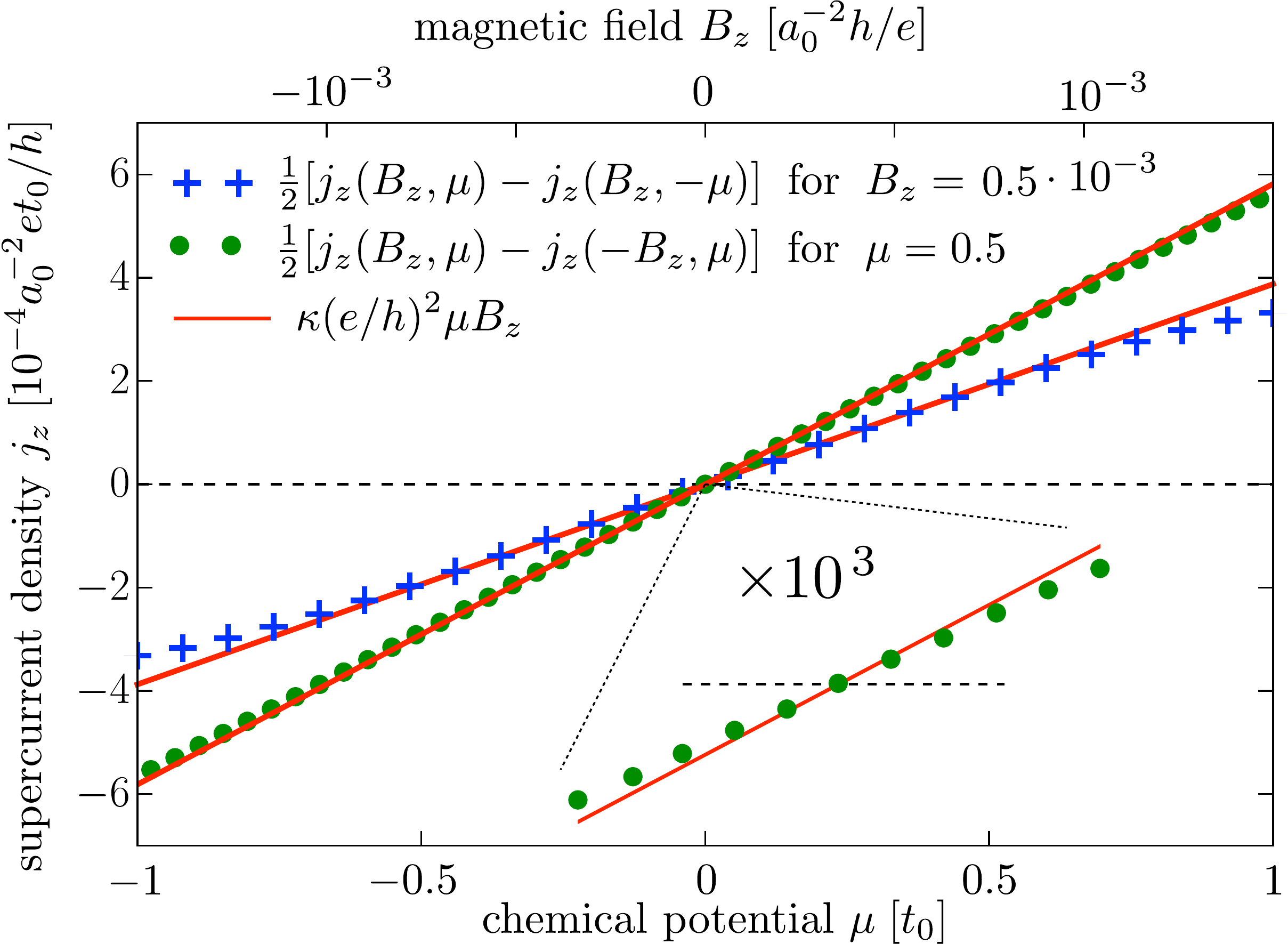}}
\caption{Data points: numerical calculation of the equilibrium supercurrent in the flux-biased circuit of Fig. \ref{fig_layout}. The parameters are $m_0=0$, $\Delta_0=0.6$, $\beta=0.5$, $\Lambda=0.45$, $W=100$, $k_{\rm B}T=0.01$; the green data points are for a fixed $\mu$ with variation of $B_z$ and the blue points for a fixed $B_z$ with variation of $\mu$. The data is antisymmetrized as indicated, to eliminate the background supercurrent from the flux bias. The solid curves are the analytical prediction \eqref{jybulk}, with $\kappa =0.775$ following directly from Eq.\ \eqref{kappadef} (no fit parameters). The $B_z$-dependent data is also shown with a zoom-in to very small magnetic fields, down to $10^{-7} a_0^{-2} h/e$, to demonstrate that the linear $B_z$-dependence continues when $l_m>W$.
}
\label{fig_parallelfield}
\end{figure}

\emph{Numerical simulation ---}
We have tested these analytical considerations in a numerical simulation of the model Hamiltonian \eqref{HBdGdef}, in the slab geometry of Fig.\ \ref{fig_layout}. At temperature $T$ the equilibrium current is given by \cite{Tin04}
\begin{equation}
I_{z}=\frac{1}{2}\sum_{n,m}\int\frac{dk_z}{2\pi}
\tanh \left(\frac{E_{nm}}{2k_{\rm B}T}\right)\Theta(E_{nm})\frac{\partial E_{nm}}{\partial A_z},\label{Izdef}
\end{equation}
where $\Theta(E)$ is the unit step function and the prefactor of $1/2$ takes care of a double counting in the BdG Hamiltonian $H$. The eigenvalues $E_{nm}(k_z)$ of $H$ are labeled by a pair of mode indices $n,m$ for motion in the $x$--$y$ plane transverse to the current. In Fig.\ \ref{fig_parallelfield} we show results for the current density $j_z=I_z/WW'$ in the $T=0$ limit, including a small thermal broadening in the numerics to improve the stability of the calculation. 

\begin{figure}[tb]
\centerline{\includegraphics[width=1\linewidth]{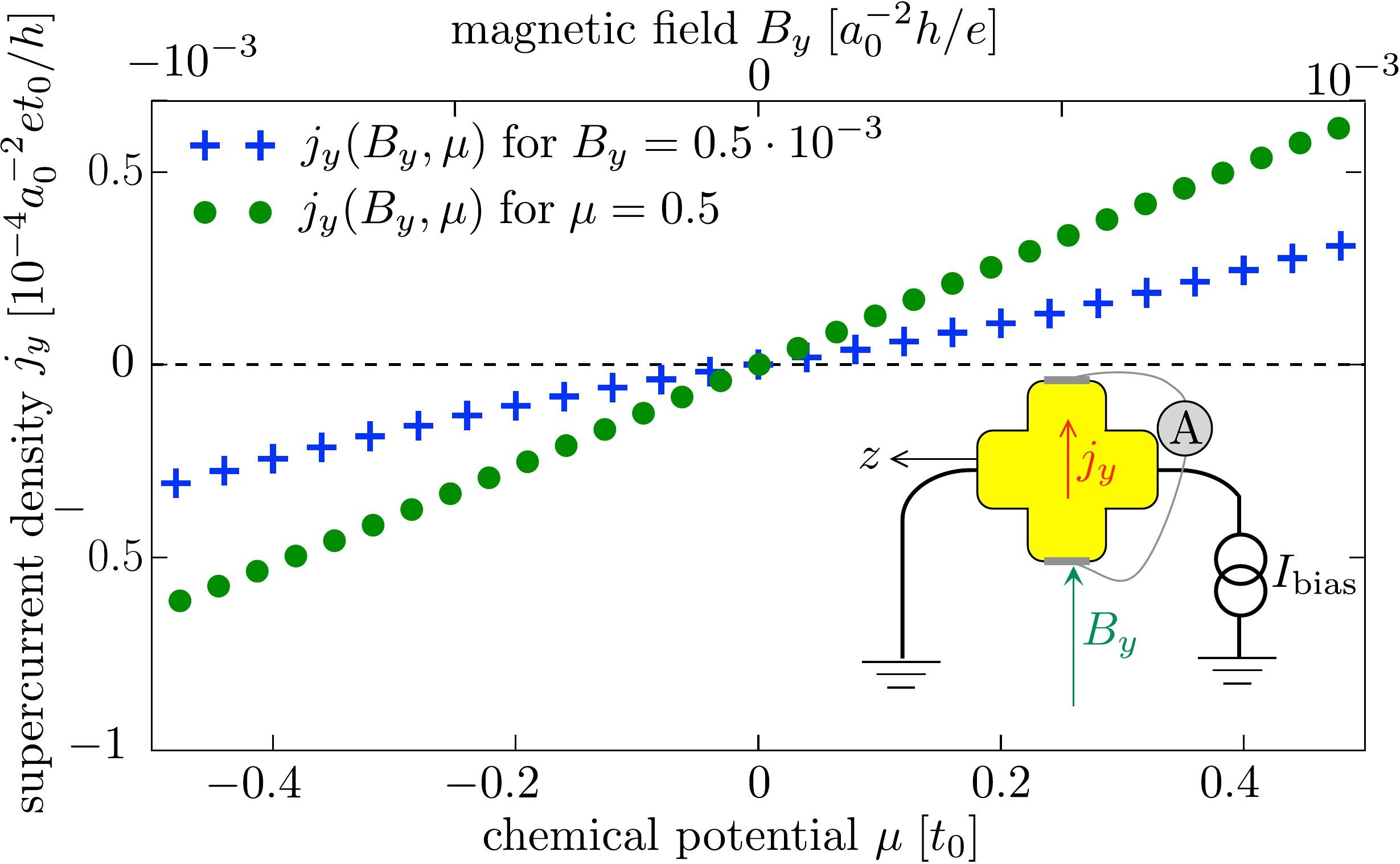}}
\caption{Same as Fig.\ \ref{fig_parallelfield} in the current-biased circuit show in the inset. No antisymmetrization of the data is needed because the measured current is perpendicular to the current bias.
}
\label{fig_perpendicularfield}
\end{figure}

We see that the numerical data is well described by the analytical result \eqref{jybulk}, with charge renormalization factor $\kappa=0.775$ from Eq.\ \eqref{kappadef}. That analytical formula was derived upon linearization in $\bm{k}$ and $\bm{A}$. A more accurate calculation \cite{supplemental} that includes the nonlinear terms in the BdG Hamiltonian gives $\kappa=0.750$, so the simple formula \eqref{kappadef} is quite accurate.

\emph{Extensions ---}
We mention extension of our findings that may help to observe the equilibrium CME in an experiment. A first extension is to smaller flux biases in the ${\cal N}=2$ regime, when two pairs of charge-conjugate cones remain gapless. The supercurrent is then given by
\begin{equation}
\frac{\partial j_z}{\partial B_z}=(\kappa_+-\kappa_-)\frac{e^2}{h^2}\mu,\;\;\kappa_\pm=\sqrt{1-\Delta_0^2/(\beta\pm\Lambda)^2},
\end{equation}
so the CME can be observed without fully gapping out one pair of cones. 

A second extension is to a current-biased, rather than flux-biased circuit, with the applied magnetic field $B_y$ perpendicular to the current bias $j_0$ in the $z$-direction. The current bias then drives the Weyl superconductor into the ${\cal N}=1$ phase via the vector potential component $A_z=\mu_0\lambda^2 j_0\equiv\Lambda/e$, with $\lambda$ the London penetration depth \cite{Tin04}. The analytical theory for this alternative configuration is more complicated, and not given here, but numerical results are shown in Fig.\ \ref{fig_perpendicularfield}. While the effect is smaller than in the flux-biased configuration, it is not superimposed on a large background supercurrent so it might be more easily observed.

A third extension concerns the inclusion of disorder. Our analysis is simplified by the assumption of a clean slab, without disorder. We expect that the chirality of the zeroth Landau level will protect the equilibrium CME from degradation by impurity scattering, in much the same way as the nonequilibrium CME is protected.

\emph{Conclusion ---}
We have shown how the chiral anomaly of an unpaired Weyl cone can be accessed in equilibrium in a superconducting Weyl semimetal. A flux bias drives the system in a state with a single charge-conjugate pair of Weyl cones, that responds to an applied magnetic field as a single species of Weyl fermions. The cancellation of the chiral magnetic effect (CME) for left-handed and right-handed Weyl fermions is removed, resulting in an equilibrium current along the field lines. The predicted size of the induced current is the same as that of the nonequilibrium CME, up to a charge renormalization of order unity, and since that dynamical effect has been observed {\cite{Kim13,Xio15,Hua15,Li15,Li16} the static counterpart should be observable as well ---  perhaps even more easily because decoherence and relaxation play no role.

In closing we note that the chiral anomaly in a crystal was originally proposed \cite{Nie83} as a condensed matter realization of an effect from relativistic quantum mechanics, and has since been an inspiration in particle physics and cosmology \cite{Boy12,Gos12,Mir15,Sto16}. The doorway to single-cone physics that we have opened here might well play a similar role.

\emph{Acknowledgments ---}
We have benefited from discussions with P. Baireuther, V. Cheianov, C. Sa\c{c}l{\i}o\u{g}lu, and B. Tarasinski. This research was supported by the Netherlands Organization for Scientific Research (NWO/OCW), an ERC Synergy Grant, by funds of the Erdal \.{I}n\"{o}n\"{u} chair, and by the T\"{U}B\.{I}TAK grant No.\ 114F163.

\clearpage
\appendix

\section{Charge renormalization in a superconducting Weyl cone}
\label{app_bulk}

We develop an effective low-energy description of the BdG Hamiltonian \eqref{HBdGdef}, to determine the charge renormalization factors that govern the equilibrium CME. In the main text we gave a simplified description, linearized in $\bm{k}$ and $\bm{A}$, valid if the Weyl points are near the center of the Brillouin zone.  Here we retain the nonlinear terms to obtain more accurate expressions valid throughout the Brillouin zone. As it turns out, our final result \eqref{kappaexact} for the charge renormalization factor is within a few percent of the simple formula \eqref{kappadef} for the parameters in the simulation of Fig.\ \ref{fig_parallelfield}.

In this Appendix \ref{app_bulk} we focus on the bulk spectrum, the surface states are considered in the next Appendix \ref{app_surface}.

\subsection{Block diagonalization}

For a real pair potential $\Delta_0$ and including the flux bias $A_z=\Lambda/e$ by the substitution $k_z\mapsto k_z-\Lambda\nu_z$, the BdG Hamiltonian is
\begin{align}
H={}&\nu_z\tau_z(\sigma_x\sin k_x+\sigma_y\sin k_y+\sigma_z\sin k_z\cos\Lambda)\nonumber\\
&+ m_{\bm k}\nu_z\tau_x\sigma_0-\mu\nu_z\tau_0\sigma_0+\beta\nu_0\tau_0\sigma_z+\Delta_0\nu_x\tau_0\sigma_0\nonumber\\
&-\nu_0\tau_z\sigma_z\cos k_z\sin\Lambda-\nu_0\tau_x\sigma_0\sin k_z\sin\Lambda,\label{HBdGstrainapp}\\
m_{\bm k}={}&m_0+\big(3-\cos k_x-\cos k_y-\cos k_z\cos\Lambda).\label{mkdefapp}
\end{align}
The $8\times 8$ matrix $H$ is constructed from the tensor product $\nu_\alpha\tau_\beta\sigma_\gamma\equiv \nu_\alpha\otimes\tau_\beta\otimes\sigma_\gamma$ of the Pauli matrices $\nu_\alpha$, $\tau_\beta$, $\sigma_\gamma$, acting respectively on the particle-hole, orbital, and spin degree of freedom.

Adapting the block-diagonalization procedure of Ref.\ \onlinecite{Bai16}, we carry out a sequence of $k_z$-dependent unitary transformations,
\begin{subequations}
\label{U1U2U3}
\begin{align}
&\tilde{H}= U_3^\dagger U_2^\dagger U_1^\dagger HU_1^{\vphantom{\dagger}}U_2^{\vphantom{\dagger}}U_3^{\vphantom{\dagger}},\\
&U_1=\exp\left(-\tfrac{1}{2}ik_z\nu_0\tau_y\sigma_z\right),\;\;U_2=\exp\left(\tfrac{1}{2}i\theta\nu_y\tau_z\sigma_z\right),\nonumber\\
& U_3=\exp\left(\tfrac{1}{2}i(\phi_0\nu_0+\phi_z\nu_z)\tau_y\sigma_z\right),
\end{align}
\end{subequations}
where the angles $\theta,\phi_0,\phi_z$ are determined by
\begin{subequations}
\begin{align}
&\cos\theta=\frac{u_k}{M_0},\;\;\sin\theta=\frac{\Delta_0}{M_0},\\
&\cos(\phi_0\pm\phi_z)=\frac{M_0\pm\sin\Lambda}{M_\pm},\\
&\sin(\phi_0\pm\phi_z)=\frac{v_k}{M_\pm},\\
&u_k=  -m_{\bm k}\sin k_z -\sin k_z \cos k_z\cos\Lambda,\\
&v_k=m_{\bm k}\cos k_z-\sin^2 k_z\cos\Lambda  ,\\
&M_0=\sqrt{\Delta_0^2+u_k^2},\\
&M_\pm=\sqrt{(M_0\pm \sin\Lambda)^2+v_k^2}.
\end{align}
\end{subequations}

We thus arrive at a transformed Hamiltonian,
\begin{align}
\tilde{H}={}&\nu_z \tau_z (\sigma_x \sin k_x + \sigma_y \sin k_y)+ \beta \nu_0 \tau_0 \sigma_z   \nonumber\\
&-\nu_z \tau_z \sigma_z \sqrt{(M_0+\nu_z\sin \Lambda)^2+v_k^2}- \mu \cos\theta  \nu_z \tau_0 \sigma_0\nonumber\\
&- \mu\sin\theta  \cos\phi_0  \nu_x \tau_z \sigma_z - \mu\sin\theta \sin\phi_0\nu_x \tau_x \sigma_0 ,\label{Htransformedstrain}
\end{align}
that for small $\mu$ is predominantly block-diagonal in the $\nu$ and $\tau$ degree of freedom.

We focus on the parameter range $M_-<\beta<M_+$ where two of the four Weyl cones are gapped by the phase bias $\Lambda$, leaving one gapless particle-hole conjugate pair. The effective low-energy Hamiltonian $H_{\rm eff}$ is then obtained by projecting $\tilde{H}$ onto the $\nu_z=-1$, $\tau_z=-1$ band,
\begin{equation}
H_{\rm eff}=\sigma_x \sin k_x + \sigma_y \sin k_y+ (\beta  - M_-)\sigma_z + \mu   \sigma_0\cos\theta.\label{Hlowenergy}
\end{equation}

The two Weyl points are at the momenta $\pm\bm{K}=(0,0,\pm K_z)$ where $M_-=\beta$. Near one of the Weyl points, to first order in $\delta\bm{k}=\bm{k}-\bm{K}$, the effective Hamiltonian represents an anisotropic Weyl cone:
\begin{equation}
H_{\bm{K}}=\sum_\alpha v_\alpha \delta k_\alpha \sigma_\alpha+\mu\sigma_0\cos\theta,\label{HeffnearWeyl}
\end{equation}
with effective velocity $\bm{v}=(1,1,-\partial M_-/\partial k_z)$ evaluated at $\bm{k}=\bm{K}$.

\subsection{Current and charge operators}

The electrical current operator
\begin{equation}
\bm{j}=-\lim_{\bm{a}\rightarrow 0}\frac{\partial}{\partial\bm{a}}H(\bm{k}-e\nu_z\bm{a})
\end{equation}
associated with the BdG Hamiltonian \eqref{HBdGstrainapp} has components
\begin{subequations}
\label{jxjyjz}
\begin{align}
j_x={}&e\nu_0\tau_z\sigma_x \cos k_x+e\nu_0\tau_x\sigma_0 \sin k_x,\\
j_y={}&e\nu_0\tau_z\sigma_y \cos k_y+e\nu_0\tau_x\sigma_0 \sin k_y,\\
j_z={}&e\nu_0\tau_z\sigma_z \cos k_z\cos\Lambda+e\nu_0\tau_x\sigma_0 \sin k_z\cos\Lambda\nonumber\\
&+e\nu_z\tau_z\sigma_z \sin k_z\sin\Lambda-e\nu_z\tau_x\sigma_0 \cos k_z\sin\Lambda.
\end{align}
\end{subequations}
The unitary transformation \eqref{U1U2U3} maps this into
\begin{equation}
\tilde{\textit{\j}}_\alpha= U_3^\dagger U_2^\dagger U_1^\dagger j_\alpha U_1^{\vphantom{\dagger}}U_2^{\vphantom{\dagger}}U_3^{\vphantom{\dagger}},
\end{equation}
resulting in
\begin{widetext}
\begin{subequations}
\begin{align}
\tilde{\textit{\j}}_x={}&e\nu_0\tau_z\sigma_x\cos k_x\cos\theta\nonumber\\
&-e\nu_0\tau_z\sigma_z\sin k_x\bigl[\cos k_z\cos\theta\sin(\phi_0+\nu_z\phi_z)-\sin k_z\cos(\phi_0+\nu_z\phi_z)\bigr]\nonumber\\
&+e\nu_0\tau_x\sigma_0\sin k_x\bigl[\cos k_z\cos\theta\cos(\phi_0+\nu_z\phi_z)+\sin k_z\sin(\phi_0+\nu_z\phi_z)\bigr]\nonumber\\
&-e\nu_x\tau_0\sigma_0\sin k_x\cos k_z\sin\theta\sin\phi_z+e\nu_y\tau_0\sigma_y\cos k_x\sin\theta\cos\phi_0\nonumber\\
&+e\nu_y\tau_y\sigma_z\sin k_x\cos k_z\sin\theta\cos\phi_z-e\nu_y\tau_y\sigma_x\cos k_x\sin\theta\sin\phi_0,\\
\tilde{\textit{\j}}_y={}&e\nu_0\tau_z\sigma_y\cos k_y\cos\theta\nonumber\\
&-e\nu_0\tau_z\sigma_z\sin k_y\bigl[\cos k_z\cos\theta\sin(\phi_0+\nu_z\phi_z)-\sin k_z\cos(\phi_0+\nu_z\phi_z)\bigr]\nonumber\\
&+e\nu_0\tau_x\sigma_0\sin k_y\bigl[\cos k_z\cos\theta\cos(\phi_0+\nu_z\phi_z)+\sin k_z\sin(\phi_0+\nu_z\phi_z)\bigr]\nonumber\\
&-e\nu_x\tau_0\sigma_0\sin k_y\cos k_z\sin\theta\sin\phi_z-e\nu_y\tau_0\sigma_x\cos k_y\sin\theta\cos\phi_0\nonumber\\
&+e\nu_y\tau_y\sigma_z\sin k_y\cos k_z\sin\theta\cos\phi_z-e\nu_y\tau_y\sigma_y\cos k_y\sin\theta\sin\phi_0,\\
\tilde{\textit{\j}}_z={}&e\nu_0\tau_z\sigma_z\cos(\Lambda+\phi_0+\nu_z\phi_z)+e\nu_0\tau_x\sigma_0\sin(\Lambda+\phi_0+\nu_z\phi_z).\label{tildejzapp}
\end{align}
\end{subequations}
\end{widetext}

Upon projection onto the $\nu_z=-1$, $\tau_z=-1$ band we thus arrive at
\begin{subequations}
\label{Jtransformed}
\begin{align}
\tilde{\textit{\j}}_x={}&e\sigma_z\sin k_x(\cos k_z\cos\theta\sin\phi_--\sin k_z\cos\phi_-)\nonumber\\
&-e\sigma_x\cos k_x\cos\theta\\
\tilde{\textit{\j}}_y={}&e\sigma_z\sin k_y(\cos k_z\cos\theta\sin\phi_--\sin k_z\cos\phi_-)\nonumber\\
&-e\sigma_y\cos k_y\cos\theta,\\
\tilde{\textit{\j}}_z={}&-e\sigma_z\cos(\Lambda+\phi_-)=e\sigma_z\partial M_-/\partial \Lambda.
\end{align}
\end{subequations}
We have abbreviated $\phi_-\equiv\phi_0-\phi_z$.

The corresponding charge operator is simply
\begin{equation}
Q=-e\partial H_{\rm eff}/\partial\mu=-e\sigma_0\cos\theta,\label{Qoperator}
\end{equation}
resulting in a charge expectation value
\begin{equation}
\label{Qaverageapp}
\begin{split}
\langle Q\rangle&=-e\cos\theta\\
&=\frac{e(3+m_0-\cos k_x-\cos k_y)\sin k_z}{\sqrt{\Delta_0^2+(3+m_0-\cos k_x-\cos k_y)\sin^2 k_z}}
\end{split}
\end{equation}
of the gapless quasiparticles. The charge changes sign as we move from one Weyl cone at $\bm{K}$ to its particle-hole conjugate at $-\bm{K}$. 

Notice that $\langle Q\rangle$ is independent of $A_z=e\Lambda$. We will make us of this later on to explain why the off-shell contributions to the CME can be neglected [see Eq.\ \eqref{Joffshell}].

\subsection{Effective Hamiltonian in the zeroth Landau level}

To apply the effective low-energy Hamiltonian \eqref{Hlowenergy} to the zeroth Landau level we include the vector potential $\bm{A}$ from an applied magnetic field to first order,
\begin{align}
H_{\rm eff}(\bm{A})={}&\sigma_x \sin k_x + \sigma_y \sin k_y+ (\beta  - M_-)\sigma_z\nonumber\\
& + \mu   \sigma_0\cos\theta-\sum_\alpha \tilde{\textit{\j}}_\alpha A_\alpha.\label{HeffA}
\end{align}
We take the vector potential $\bm{A}=(0,B_z x,0)$ for a magnetic field $B_z$ in the $z$-direction and linearize with respect to $k_x$. This linearization also eliminates $k_x$ from the mass term $m_{\bm{k}}$, which would otherwise interfere with the $x$-dependent $\bm{A}$ when we perform the unitary transformations \eqref{U1U2U3}. We thus obtain
\begin{subequations}
\begin{align}
H_{\rm eff}={}&\sigma_x k_x + \sigma_y \sin k_y+ (\beta  - M_-)\sigma_z\nonumber\\
& + \mu   \sigma_0\cos\theta-eB_z x (V_y\sigma_y+V_z\sigma_z),\\
V_y={}&-\cos k_y\cos\theta,\\
V_z={}&\sin k_y(\cos k_z\cos\theta\sin\phi_- -\sin k_z\cos\phi_- ).
\end{align}
\end{subequations}

The $x$ and $k_x=-i\partial/\partial x$ dependent parts of the Hamiltonian govern the decay of the wave function when $x\rightarrow\pm\infty$, according to
\begin{subequations}
\begin{align}
&\partial\psi/\partial x=i\sigma_x eB_z x(V_y\sigma_y+V_z\sigma_z)\psi,\nonumber\\
&\Rightarrow\psi(x)\propto\exp\left(-\tfrac{1}{2}eB_z x^2\sqrt{V_y^2+V_z^2}\,\right)|V\rangle,\\
&|V\rangle=\begin{pmatrix}
V_y+\sqrt{V_y^2+V_z^2}\\
-iV_z
\end{pmatrix}.
\end{align}
\end{subequations}

The energy $E_0(k_y,k_z)$ of the zeroth Landau level then follows upon projection of $H_{\rm eff}$ onto $|V\rangle$,
\begin{align}
E_0(k_y,k_z)={}&\frac{\langle V|H_{\rm eff}|V\rangle}{\langle V|V\rangle}\nonumber\\
={}&\frac{(\beta -M_- )V_y-V_z\sin k_y}{\sqrt{V_y^2+V_z^2}}+\mu\cos\theta.\label{LLLanalytics}
\end{align}

Near each of the two Weyl points at $\bm{k}=(0,0,\pm K_z)+\delta\bm{k}$ this reduces to the dispersion
\begin{equation}
\begin{split}
&E_\pm(k_z)=v_0 \delta k_z- q_\pm \mu+{\cal O}(\delta k^2),\\
&q_\pm=-\cos\theta\bigr|_{K_z},\;\;v_0=-\left.\frac{\partial M_-}{\partial k_z}\right|_{K_z}.
\end{split}
\end{equation}
of a zeroth Landau level that propagates chirally (unidirectionally) in the $z$-direction with the same velocity $v_0$ and opposite charge $q_\pm$.

\begin{figure}[tb]
\centerline{\includegraphics[width=0.7\linewidth]{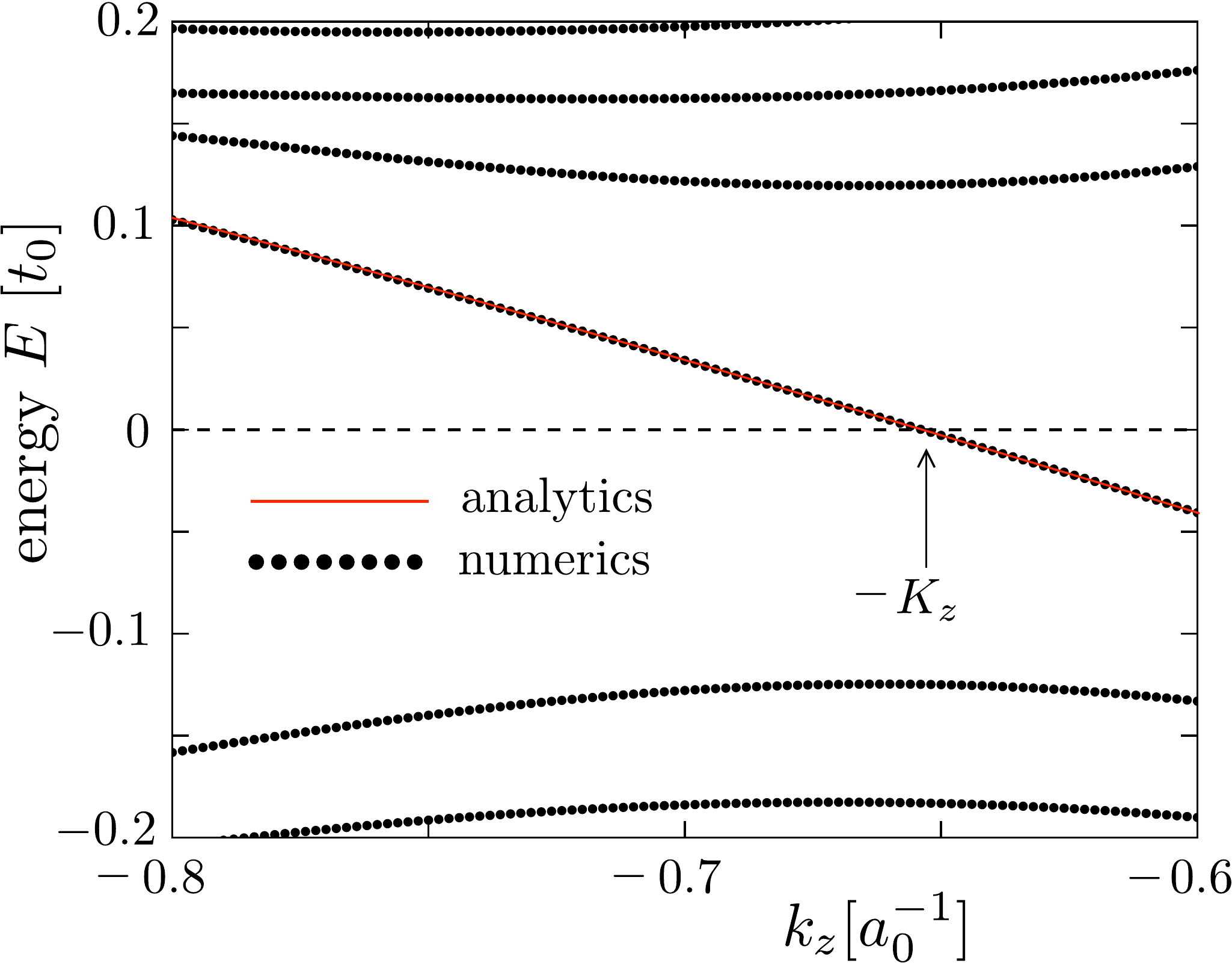}}
\caption{
Data points: Numerical results for the band structure of the Weyl superconductor near the hole-like Weyl point at $-K_z$, showing the first few Landau levels in a magnetic field $B_z=5\cdot 10^{-4}\,a_0^{-2}h/e$ (other parameters $m_0=0$, $\Delta_0=0.6$, $\beta=0.5$, $\Lambda=0.45$, $W=50$, $k_y=0$, $\mu=0$). Red curve: Analytical result \eqref{LLLanalytics} in the chiral zeroth Landau level, plotted without any fit parameters.
}
\label{fig:energykz}
\end{figure}

In Fig.\ \ref{fig:energykz} we compare the dispersion \eqref{LLLanalytics} in the zeroth Landau level, derived from the effective low-energy Hamiltonian \eqref{HeffA}, with the numerical result from the full Hamiltonian \eqref{HBdGstrainapp}. The agreement is very good without any adjustable parameters, giving confidence in the reliability of the low-energy description. 

\subsection{Renormalized charge for the CME}

To make contact with the single-cone Hamiltonian \eqref{tildecalHdef} from the main text, we seek the charge and velocity renormalization near the Weyl point at $\bm{K}$. The current and charge operators \eqref{Jtransformed} and \eqref{Qoperator} enter into the effective Hamiltonian \eqref{HeffnearWeyl} as
\begin{subequations}
\label{HeffnearWeylwithA}
\begin{align}
&H_{\bm{K}}=\sum_\alpha v_\alpha (\delta k_\alpha-q_\alpha A_\alpha) \sigma_\alpha-q_0\mu\sigma_0,\\
&\left.
\begin{array}{l}
\bm{v}=(1,1,-\partial M_-/\partial k_z),\\
q_0=-\cos\theta,\\
\bm{q}=-e\left(\cos\theta,\cos\theta,1/\cos\theta\right),
\end{array}
\right\}
\text{at}\;\;\;\bm{k}=\bm{K}.
\end{align}
\end{subequations}
We have linearized in the momentum $\delta{\bm k}=\bm{k}-\bm{K}$ and vector potential $\bm{A}$ and we have used the fact that
\begin{equation}
\displaystyle{\frac{\partial M_-/\partial \Lambda}{\partial M_-/\partial k_z}}=\frac{1}{\cos\theta}.
\end{equation}

From Eq.\ \eqref{jybulk} we find the contribution from the zeroth Landau level to the equilibrium supercurrent density,
\begin{equation}
\begin{split}
&\frac{\partial j_z}{\partial B_z}=\frac{q_0 q_y q_z}{h^2}\mu=\kappa\frac{e^2}{h^2}\mu,\\
&\kappa=-\cos\theta\bigr|_{\bm{k}=\bm{K}}=\frac{(1+m_0)\sin K_z}{\sqrt{\Delta_0^2+(1+m_0)^2\sin^2 K_z}},
\end{split}
\label{kappaexact}
\end{equation}
with $\bm{K}=(0,0,K_z)$ determined by the equation $M_-=\beta$.

\begin{figure}[tb]
\centerline{\includegraphics[width=0.6\linewidth]{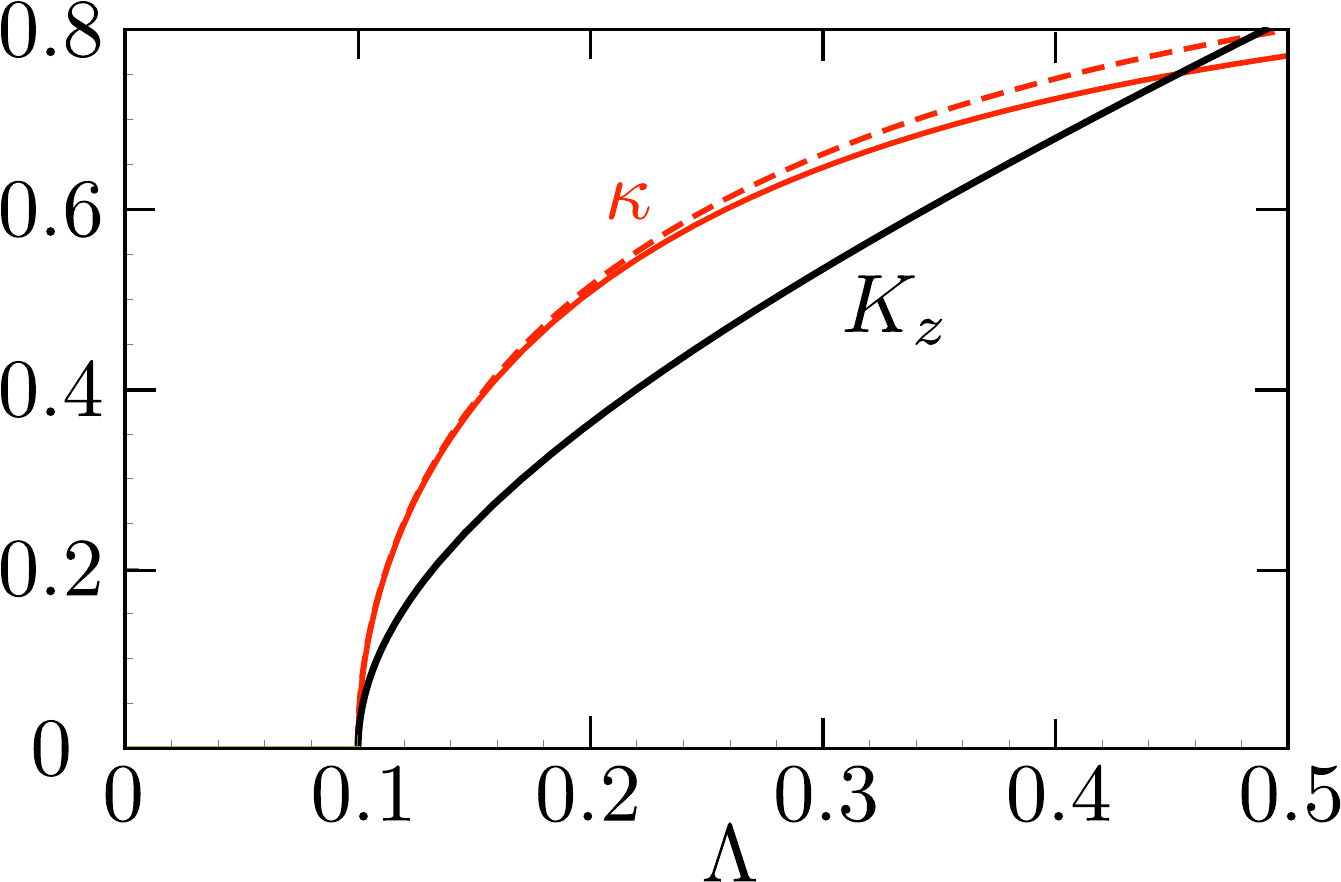}}
\caption{
Black curve: Momentum $K_z$ of the Weyl point as a function of the flux bias $\Lambda$, calculated from the solution of $M_-=\beta$ for the parameters $m_0=0$, $\Delta_0=0.6$, $\beta=0.5$. Red curves: The corresponding charge renormalization factor $\kappa$, from Eq.\ \eqref{kappaexact} (solid curve) and from the small-$K_z$ approximation \eqref{kappadef} (dashed curve). The curves terminate at the value $\Lambda=\Delta_0-\beta=0.1$ where a gap opens in the Weyl cone and the solution to $M_-=\beta$ becomes imaginary.
}
\label{fig:kappa}
\end{figure}

For the parameter values of Fig.\ \ref{fig_parallelfield} we find $K_z=0.747$, resulting in the charge renormalization factor $\kappa=0.750$. The formula \eqref{kappadef} from the linearized theory in the main text gives $\kappa=0.775$ for the same parameter values. It is remarkable how accurate that simple formula is, see Fig.\ \ref{fig:kappa}, even when $K_z$ is not much smaller than unity.

\section{Surface Fermi arc}
\label{app_surface}

In App.\ \ref{app_bulk} we gave a low-energy description of the bulk Weyl cones. We now turn to the surface states, to derive the dispersion relation shown in Fig.\ \ref{fig:surface_band} of the main text and to demonstrate that the Fermi arc carries an approximately neutral current along the surface.

\subsection{Boundary condition}

In the slab geometry of Fig.\ \ref{fig_layout} the Weyl superconductor is confined to the inner region $|x|<W/2$ by an infinite mass in the outer region $|x|>W/2$. The requirement of a decaying wave function in the outer region, where $m_0\rightarrow\infty$, implies that the wave function at the interfaces satisfies 
\begin{equation}
(1\pm \nu_0\tau_y\sigma_x)\psi(\pm W/2)=0.
\end{equation}
The unitary transformation \eqref{U1U2U3} changes this boundary condition into
\begin{widetext}
\begin{align}
&(1\pm U_b)\tilde{\psi}(\pm W/2)=0,\\
U_b={}&U_3^\dagger U_2^\dagger U_1^\dagger \nu_0\tau_y\sigma_x U_1U_2U_3\nonumber\\
={}&\nu_0\tau_0\sigma_y[\cos k_z\sin(\phi_0+\nu_z\phi_z) -\cos\theta\sin k_z\cos(\phi_0+\nu_z\phi_z)]\nonumber\\
&+\nu_0\tau_y\sigma_x[\cos k_z\cos(\phi_0+\nu_z\phi_z) +\cos\theta\sin k_z\sin(\phi_0+\nu_z\phi_z)]\nonumber\\
&+\nu_y\tau_z\sigma_x\sin\theta\sin k_z\cos\phi_z+\nu_x\tau_x\sigma_y\sin\theta\sin k_z\sin\phi_z,
\end{align}
for the transformed wave function $\tilde\psi=U_3^\dagger U_2^\dagger U_1^\dagger \psi$.
\end{widetext}

For later use we note that the two matrices $U_0=\nu_z\tau_z\sigma_y$ and $U_b$ commute, so they can be jointly diagonalized. Each matrix has eigenvalues $\pm 1$, we seek the eigenspace where both eigenvalues have same sign. The two orthonormal eigenvectors $u_1$ and $u_2$ with eigenvalue $-1$ are given by
\begin{subequations}
\label{u1u2def}
\begin{align}
u_1={}&\tfrac{1}{2}Z_0^{-2}\bigl(i Z_1,Z_1,-i Z_2,Z_2,0,0,i Z_4,Z_4\bigr),\\
u_2={}&\tfrac{1}{2}Z_0^{-2}\bigl(i \cos \phi_z Z_4,\cos \phi_z Z_4,i \sin \phi_z Z_4,\nonumber\\
&\qquad-\sin \phi_z Z_4,-iZ_0,Z_0,-i Z_3,-Z_3\bigr),\\
Z_0={}&1-\cos k_z\sin\phi_- + \sin k_z\cos\theta\cos\phi_-,\\
Z_1={}&\sin \phi_0 \sin k_z \cos \theta+\cos \phi_0 \cos k_z+\sin \phi_z,\\
Z_2={}&\cos \phi_0 \sin k_z \cos \theta-\sin \phi_0 \cos k_z+\cos \phi_z,\\
Z_3={}&\cos k_z \cos \phi_- +\sin k_z\cos \theta \sin \phi_-,\\
Z_4={}&\sin k_z \sin \theta.
\end{align}
\end{subequations}
The eigenspace with eigenvalue $+1$ of $U_0$ and $U_b$ is spanned by $u_3=\nu_0\tau_0\sigma_z u_1$ and $u_4=\nu_0\tau_0\sigma_z u_2$.

\subsection{Construction of the surface state}

For $M_-<\beta<M_+$ there is only one pair of gapless Weyl cones, so there is a single low-energy surface state connecting them. We assume that $W$ is sufficiently large that we can treat the two surfaces at $x=\pm W/2$ independently. Let us consider the surface state $\tilde{\psi}$ at $x=W/2$. It should be a solution of $\tilde{H}\tilde{\psi}=E\tilde{\psi}$ that decays for $x\rightarrow -\infty$ and that satisfies the boundary condition $U_b\tilde{\psi}=-\tilde{\psi}$ at $x=W/2$.

We first solve this matching problem to zeroth order in $\mu$, when the Hamiltonian \eqref{Htransformedstrain} reduces to
\begin{align}
\tilde{H}_0={}&\nu_z \tau_z (\sigma_x \sin k_x + \sigma_y \sin k_y)+ \beta \nu_0 \tau_0 \sigma_z   \nonumber\\
&-\nu_z \tau_z \sigma_z \sqrt{(M_0+\nu_z\sin \Lambda)^2+v_k^2}.\label{Htransformedstrain2}
\end{align}
We linearize in $k_x=-i\partial /\partial x$ and obtain the solution of $\tilde{H}_0\tilde{\psi}=E_0\tilde{\psi}$ in the form
\begin{align}
\tilde{\psi}(x)={}&\exp\biggl[i\delta x\,\nu_z\tau_z\sigma_x\bigl(E_0-\nu_z\tau_z\sigma_y\sin k_y- \beta \nu_0 \tau_0 \sigma_z\nonumber\\
&+\nu_z \tau_z \sigma_z \sqrt{(M_0+\nu_z\sin \Lambda)^2+v_k^2}\bigr)\biggr]\tilde{\psi}(W/2),\label{tildepsisolution}
\end{align}
abbreviating $\delta x=x-W/2$.

For $E_0=-\sin k_y$ the solution \eqref{tildepsisolution} that decays for $\delta x\rightarrow-\infty$ is an eigenvector of $U_0=\nu_z\tau_z\sigma_y$ with eigenvalue $-1$:
\begin{align}
\tilde{\psi}(x)={}&\biggl(0,0,-i C_1 e^{(\beta+M_+ )\delta x},C_1 e^{(\beta+M_+ )\delta x},\nonumber\\
&-i C_2 e^{(\beta+M_-)\delta x)},C_2 e^{(\beta+M_-)\delta x},\nonumber\\
&i C_3 e^{(\beta-M_-)\delta x},C_3 e^{(\beta-M_-)\delta x}\biggr).\label{tildepsiresult}
\end{align}
To satisfy the boundary condition at $x=W/2$, the coefficients $C_1$, $C_2$, $C_3$ should be chosen such that $\tilde{\psi}(W/2)=(0,0,-iC_1,C_1,-iC_2,C_2,iC_3,C_3)$ is a superposition of the eigenvectors $u_1$ and $u_2$ in Eq.\ \eqref{u1u2def}. This results in
\begin{equation}
\begin{split}
&C_1=Z_1 Z_4\sin\phi_z+Z_2 Z_4\cos\phi_z,\;\;C_2=-Z_0 Z_1,\\
&C_3=Z_1 Z_3+Z_4^2\cos\phi_z,
\end{split}
\end{equation}
up to an overall normalization constant.

\subsection{Surface dispersion relation}

We now add to the zeroth order energy $E_0=-\sin k_y$ the contribution $\delta E_\mu$ from the chemical potential in first order perturbation theory,
\begin{align}
\delta E_\mu={}&\frac{\langle\tilde\psi|\delta \tilde{H}|\tilde\psi\rangle}{\langle\tilde\psi|\tilde\psi\rangle},\\
\delta\tilde{H}={}&\tilde{H}-\tilde{H}_0=- \mu \cos\theta  \nu_z \tau_0 \sigma_0
- \mu\sin\theta  \cos\phi_0  \nu_x \tau_z \sigma_z \nonumber\\
&\qquad\qquad- \mu\sin\theta \sin\phi_0\nu_x \tau_x \sigma_0.
\end{align}
Two of the three $\mu$-dependent terms in $\delta \tilde{H}$ mix the $\nu=\pm 1$ bands in the bulk. The small parameter that governs the $\nu$-band mixing is $\delta_{\rm mix}=(\beta-M_-)/(\beta+M_+)$. If we neglect this mixing and project both $\tilde{\psi}$ and $\tilde{H}$ onto the $\nu=-1$ band, we have simply
\begin{equation}
\delta E_\mu = \mu \cos\theta.
\end{equation}

In the same way we include to first order the contribution $\delta E_B$ from the magnetic field with vector potential $A_y=B_z x$,
\begin{equation}
\delta E_B=-B_z\frac{\langle\tilde\psi|x\tilde{\textit{\j}}_y|\tilde\psi\rangle}{\langle\tilde\psi|\tilde\psi\rangle}=-\tfrac{1}{2}WeB_z\cos k_y\cos\theta,
\end{equation}
where we have projected $\tilde{\psi}$ and $\tilde{\textit{\j}}_y$ onto the $\nu=-1$ band and taken the large-$W$ limit of the expectation value.

Collecting results we thus obtain the dispersion relation $E_{\rm surface}(k_y,k_z)$ for the surface Fermi arc,
\begin{align}
&E_{\rm surface}=-\sin k_y-(\tfrac{1}{2}WeB_z\cos k_y-\mu)\cos\theta\nonumber\\
&=-\sin k_y\nonumber\\
&\quad+\frac{(\tfrac{1}{2}WeB_z\cos k_y-\mu)(2+m_0-\cos k_y)\sin k_z}{\sqrt{\Delta_0^2+(2+m_0-\cos k_y)\sin^2 k_z}}.\label{Esurface}
\end{align}
This is for the surface at $x=W/2$. For the opposite surface at $x=-W/2$ we should substitute $k_y\mapsto -k_y$.

From Eq.\ \eqref{Esurface} we calculate the expectation values of the charge $\langle Q\rangle$ and the electrical current $\langle j_z\rangle$ of the surface state,
\begin{equation}
\begin{split}
&\langle Q\rangle=-e\frac{\partial E_{\rm surface}}{\partial \mu}=-e\cos\theta,\\
&\langle j_z\rangle=-e\frac{\partial E_{\rm surface}}{\partial \Lambda}=0,
\end{split}
\label{Qjzsurface}
\end{equation}
the same on both surfaces. The Fermi arc transports no charge in the $z$-direction --- up to corrections of order $\delta_{\rm mix}$ from the band mixing. The approximately neutral current in a Fermi arc explains why the calculation of the CME including only the chiral Landau level in the bulk agrees so well with the numerics in Fig.\ \ref{fig_perpendicularfield}.

\begin{figure}[tb]
\centerline{\includegraphics[width=0.7\linewidth]{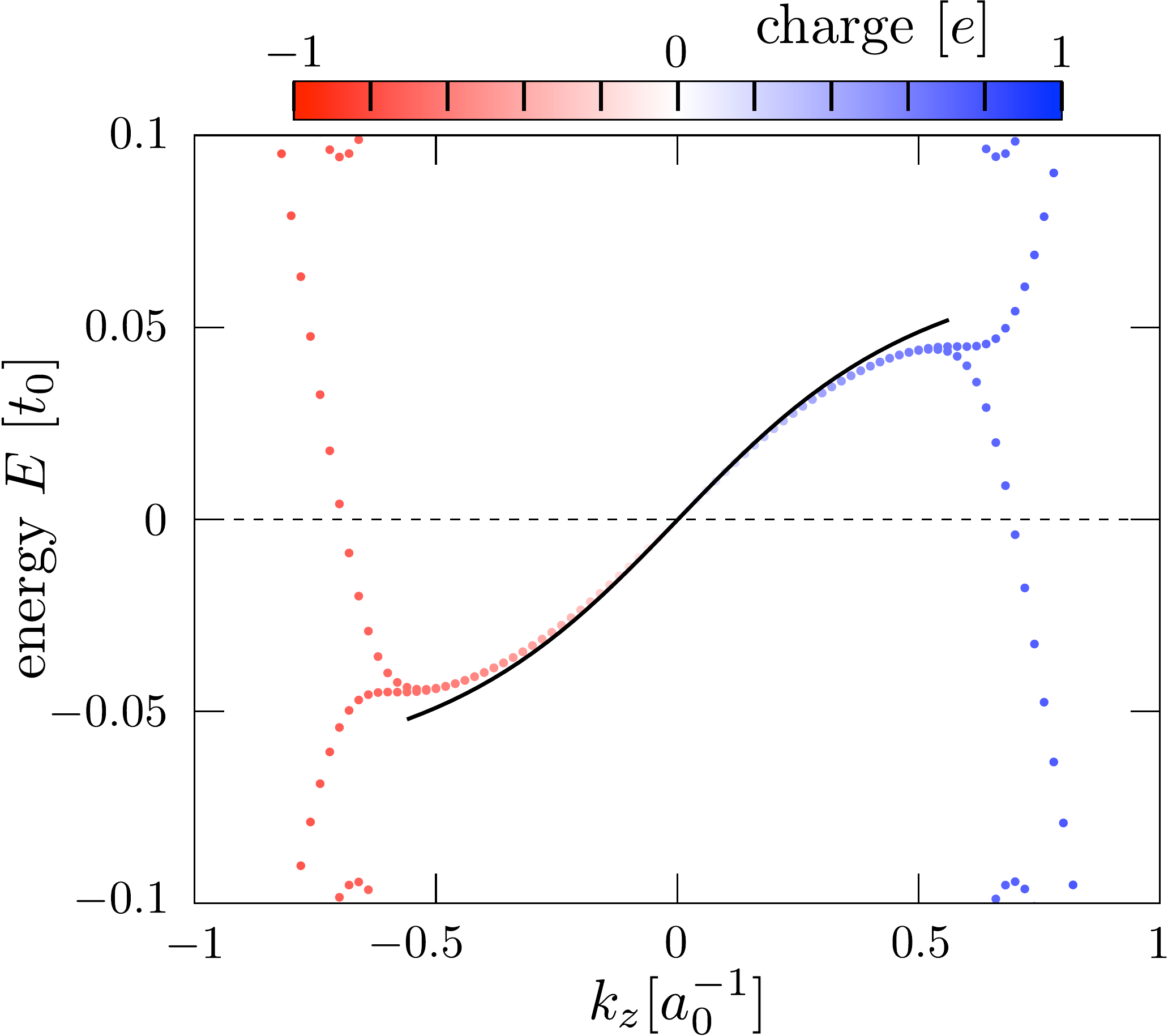}}
\caption{
Data points: Dispersion of the surface states connecting the electron-like and hole-like zeroth Landau levels, for the same parameters as Fig.\ \ref{fig:energykz}. The color scale gives the charge expectation value. The black curve is the analytical dispersion \eqref{Esurface} of the surface Fermi arc.
}
\label{fig:surfacedispersion}
\end{figure}

\begin{figure}[tb]
\centerline{\includegraphics[width=0.9\linewidth]{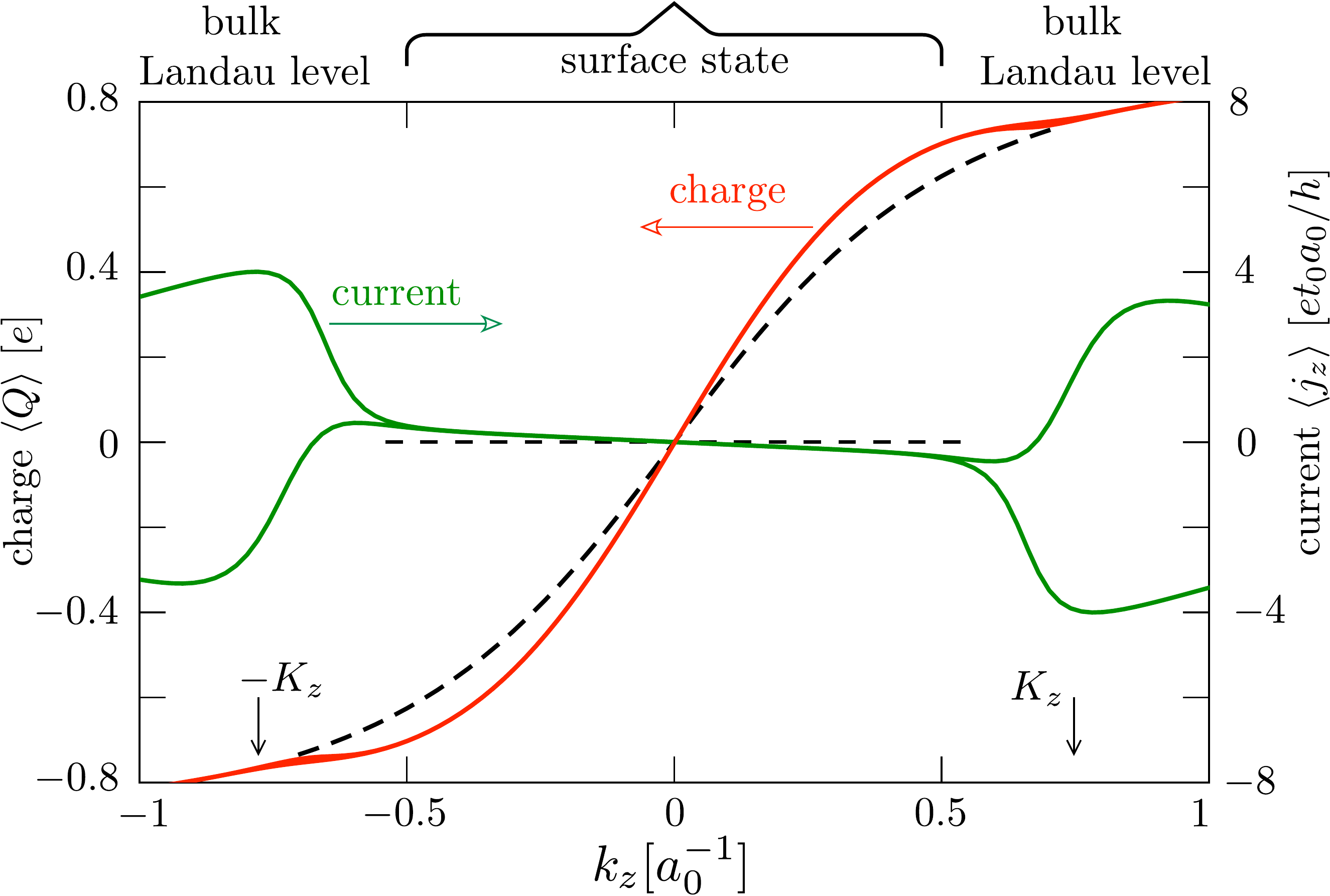}}
\caption{
Solid curves: Expectation value of charge $Q$ (red, left axis) and electrical current $j_z$ (green, right axis), for the same parameters as Fig.\ \ref{fig:surfacedispersion}. The black dashed curves are the analytical result \eqref{Qjzsurface} for the surface state. The electrical current is predominantly carried by the bulk Landau level, while the surface Fermi arc carries an approximately neutral current.
}
\label{fig:chargecurrent}
\end{figure}

In Figs.\ \ref{fig:surfacedispersion} and \ref{fig:chargecurrent} we compare these analytical results for the surface dispersion, charge, and current with the numerical data. The agreement is quite satisfactory, without any adjustable parameter.  

\section{Derivation of the renormalized-charge formula for the CME}
\label{app_CME}

Equation \eqref{jybulk} in the main text for the equilibrium CME in a superconductor has the form expected for a single Weyl cone, modified by charge renormalization. We give a derivation of this formula.

\subsection{On-shell and off-shell contributions}

The equilibrium supercurrent 
\begin{equation}
I_{z}=\frac{1}{2}\sum_{n,m}\int\frac{dk_z}{2\pi}\Theta(E)
\tanh \left(\frac{E}{2k_{\rm B}T}\right)\frac{\partial E}{\partial A_z}\label{Izdefapp}
\end{equation}
is not a Fermi surface property, but contains contributions over a range of energies $E=E_{nm}(k_z)>0$ even in the limit that the temperature $T$ goes to zero. For the CME we seek a contribution to $I_z$ that is linear in the chemical potential $\mu$, measured relative to the Weyl points. As we will now show, the derivative $\partial I/\partial \mu$ in the limit $\mu\rightarrow 0$ has predominantly Fermi-surface {\em (on-shell)} contributions, which at $T=0$ can be written as a sum over propagating modes at the Fermi energy $E=0$.

Using particle-hole symmetry (relating states at energy $\pm E$ carrying opposite current $\pm \partial E/\partial A_z$) we rewrite Eq.\ \eqref{Izdefapp} as an integral over all states of positive and negative energies,
\begin{equation}
I_{z}=-\frac{1}{2}\sum_{n,m}\int\frac{dk_z}{2\pi}\,f(E)\frac{\partial E}{\partial A_z},\label{Izdefapp2}
\end{equation}
weighted by the Fermi function 
\begin{equation}
f(E)=\left(1+e^{E/k_{\rm B}T}\right)^{-1}=\tfrac{1}{2}-\tfrac{1}{2}\tanh(E/2k_{\rm B}T).
\end{equation}

The derivative of the energy in Eq.\ \eqref{Izdefapp2} gives the expectation value of the electrical current operator $j_z= -\partial H/\partial A_z$ in the state with energy $E$,
\begin{equation}
\langle j_z\rangle_E=-\langle\partial H/\partial A_z\rangle_E=-\partial E/\partial A_z,
\end{equation}
according to the Hellmann-Feynman theorem. Two other expectation values that we need are those of the velocity operator $v_z=\partial H/\partial k_z$ and the charge operator $Q=-e\partial H/\partial \mu$, given by
\begin{equation}
\langle v_z\rangle_E=\partial E/\partial k_z,\;\;\langle Q\rangle_E=-e\partial E/\partial \mu.
\end{equation}

We take the derivative with respect to $\mu$ of Eq.\ \eqref{Izdefapp2}:
\begin{align}
\frac{\partial I_{z}}{\partial \mu}={}&{\cal J}_{\text{on-shell}}+{\cal J}_{\text{off-shell}},\label{dIalphadmu}\\
{\cal J}_{\text{on-shell}}={}&-\frac{1}{2e}\sum_{n,m}\int\frac{dk_z}{2\pi}f'(E)\langle Q\rangle_E\langle j_z\rangle_E,\label{Jonshell}\\
{\cal J}_{\text{off-shell}}={}&\frac{1}{2e}\sum_{n,m}\int\frac{dk_z}{2\pi}f(E)\frac{\partial}{\partial 
A_z}\langle Q\rangle_E.\label{Joffshell}
\end{align}
At low temperatures, when $-f'(E)\rightarrow \delta(E)$ becomes a delta function, the on-shell contribution ${\cal J}_{\text{on-shell}}$ involves only Fermi surface properties. It is helpful to rewrite it as a sum over modes at the Fermi energy. For that purpose we replace the integration over $k_z$ by an energy integration weighted with the density of states:
\begin{equation}
{\cal J}_{\text{on-shell}}=-\frac{1}{4\pi e}\sum_{n,m}\int_{-\infty}^\infty dE\, f'(E)\left|\frac{\partial E}{\partial k_z}\right|^{-1}
\langle Q\rangle_E\langle j_z\rangle_E.\label{IalphaFermileveldE}
\end{equation}

This equation may be written in a more suggestive form by defining a {\em vector charge}
\begin{equation}
\bm{Q}=(Q_x,Q_y,Q_z),\;\;\text{with}\;\;Q_\alpha(E) \equiv \frac{\langle j_\alpha\rangle_E}{\langle v_\alpha\rangle_E},
\end{equation}
which may be different from the average (scalar) charge $Q_0\equiv\langle Q\rangle_E$ because the average of the product of charge and velocity may differ from the product of the averages.(For example, the coherent superposition of a right-moving electron and a left-moving hole has zero average charge and zero average velocity, but nonzero average electrical current.) We finally arrive at
\begin{align}
{\cal J}_{\text{on-shell}}={}&-\frac{1}{4\pi e}\sum_{n,m}\int_{-\infty}^\infty dE\, f'(E)\nonumber\\
&\times Q_0(E) Q_z(E)\,\bigl({\rm sign}\,\langle v_z\rangle_E\bigr).\label{IalphaFermilevelvectorQ}
\end{align}
At zero temperature a sum over modes remains,
\begin{equation}
{\cal J}_{\text{on-shell}}=\frac{1}{2}\frac{e}{h}\sum_{n,m}
\frac{Q_0 Q_z}{e^2}\,\bigl({\rm sign}\,\langle v_z\rangle\bigr)\biggr|_{E_{nm}=0},\label{IalphaFermilevel}
\end{equation}
where we have restored the units of $\hbar=h/2\pi$. The subscript $n,m$ labels the mode indices of a propagating mode in the $z$-direction at the Fermi energy ($E=0$). 

\subsection{Application to the zeroth Landau level}

We evaluate Eq.\ \eqref{IalphaFermilevel} for the effective Hamiltonian \eqref{HeffnearWeylwithA}
in the zeroth Landau level near the Weyl point at $\bm{K}$ and its charge-conjugate at $-\bm{K}$. The two Weyl points have opposite sign of both the scalar charge $Q_0=-e\cos\theta$ and the vector charge $Q_z=-e/\cos\theta$, and the same ${\rm sign}\,\langle v_z\rangle=\chi\,({\rm sign}\,B_z)$, so their contributions add. The Landau level degeneracy is
\begin{equation}
{\cal N}=\frac{1}{h}WW' |B_z Q_{y}|=\frac{e}{h}WW'| B_z\cos\theta|,
\end{equation}
Substitution into Eq.\ \eqref{IalphaFermilevel}, times two for two Weyl points, gives the on-shell contribution to the zero-temperature equilibrium current,
\begin{equation}
{\cal J}_{\text{on-shell}}=\frac{e}{h}{\cal N}\frac{Q_0 Q_z}{e^2}\chi(\,{\rm sign}\,B_z)=WW'\frac{e^2}{h^2}\kappa \chi B_z,
\end{equation}
with charge renormalization factor
\begin{align}
\kappa=\lim_{\bm{k}\rightarrow\bm{K}}|\cos\theta|.
\end{align}

This confirms Eq.\ \eqref{jybulk} in the main text (where we took a positive chirality $\chi$), provided that we can neglect 1) contributions from the surface states; and 2) off-shell contributions from the bulk states. A numerical demonstration that these contributions can be neglected is provided in Fig.\ \ref{fig_perpendicularfield}, where the full expression \eqref{Izdefapp} is evaluated in a slab geometry. Analytical justification comes from the effective low-energy Hamiltonian, which shows that 1) $\partial E/\partial A_z=e\partial E/\partial \Lambda$ vanishes on the surface in view of Eq.\ \eqref{Qjzsurface}; and 2) $\partial \langle Q\rangle_E/\partial A_z$ vanishes in the bulk in view of Eq.\ \eqref{Qaverageapp}.


\begin{thebibliography}{99}
\bibitem{vortex} For magnetic lengths $l_m=\sqrt{\hbar/eB}$ below the thickness $W$ of the slab, the parallel magnetic field $B$ may induce vortices in the order parameter. We take a uniform order parameter in our analysis, however, the numerical data in Fig.\ \ref{fig_parallelfield} shows that our results extend down to the lowest fields with $l_m\gg W$, when vortex formation is suppressed.
\bibitem{note0} The chirality $\chi$ of a Weyl cone determines the sign of the dispersion of the zeroth Landau level in a magnetic field:  ${\rm sign}\,(dE/dk_z)=\chi\,{\rm sign}\,(B_z)$. In the flux-biased Weyl superconductor $\chi=-{\rm sign}\,\Lambda$, as one can see in Fig.\ \ref{fig:surface_band}.
\end{thebibliography}
\end{document}